\documentclass[letter,twocolumn]{jpsj2} %% two-column layout
\title{Magnetic-Field Induced Bose-Einstein Condensation of Magnons and\\
Critical Behavior in Interacting Spin Dimer System TlCuCl$_3$}

\author{Fumiko {\sc Yamada}\thanks{E-mail: yamada@lee.phys.titech.ac.jp.}, Toshio {\sc Ono}, Hidekazu {\sc Tanaka}, Gr$\acute{\rm e}$goire {\sc Misguich}$^{1}$, Masaki {\sc Oshikawa}$^{2}$ \\
and Toshiro {\sc Sakakibara}$^{2}$}

\inst{Department of Physics, Tokyo Institute of Technology, Oh-okayama, Meguro-ku, Tokyo 152-8551\\
$^{1}$Service de Physique Th$\acute{e}$orique, CEA-Saclay, 91191 Gif-sur-Yvette C$\acute{e}$dex, France\\
$^{2}$Institute for Solid State Physics, The University of Tokyo, Kashiwanoha, Kashiwa, Chiba 277-8581}

\abst{Magnetization measurements were performed to investigate the critical behavior of the field-induced magnetic ordering in gapped spin system TlCuCl$_3$. The critical density of the magnons was obtained as a function of temperature and the magnon-magnon interaction constant was evaluated. The experimental phase boundary for $T < 5$ K agrees almost perfectly with the magnon BEC theory based on the Hartree-Fock approximation with realistic dispersion relations. The phase boundary can be described by the power law $\left[H_{\rm N}(T)-H_{\rm c}\right] \propto T^{\phi}$. With decreasing fitting temperature range, the critical exponent ${\phi}$ decreases and converges at $\phi_{\rm BEC} =3/2$ predicted by the magnon BEC theory.}

\kword{TlCuCl$_3$, spin dimer system, magnons, quantum phase transition, Bose-Einstein condensation, critical behavior}

\begin{document}
\maketitle

%\section{Introduction} 
Quantum spin system composed of antiferromagnetic spin dimer often shows a gapped singlet ground state. In an external magnetic field exceeding the energy gap $\Delta$, $S_z = 1$ component of the spin triplet is created in the system. The field-induced $S_z = 1$ component has the characteristics of boson and is called magnon or triplon. Magnons move to neighboring dimers and interact with one another due to the transverse and longitudinal components of the interdimer exchange interactions, respectively. Consequently, the spin dimer system in the magnetic field can be represented as an interacting boson system\,\cite{Rice}. Magnons can undergo Bose-Einstein condensation (BEC) in a magnetic field higher than the critical field $H_{\rm c}={\Delta}/g{\mu}_{\rm B}$, which leads to field-induced magnetic ordering (FIMO)\,\cite{Nikuni,Wessel}. Nikuni {\it et al.} \cite{Nikuni} discussed the FIMO observed in TlCuCl$_3$ \cite{O_mag}, applying the Hartree-Fock (HF) approximation to a simplified model
\begin{eqnarray}
{\cal{H}} = \sum_{{\mib k}}\left(\varepsilon_{{\mib k}}-\mu\right)a^{\dag}_{{\mib k}}a_{{\mib k}}
+\frac{U}{2N}\sum_{{\mib k},{\mib k}',{\mib q}}a^{\dag}_{{\mib k}+{\mib q}}a^{\dag}_{{\mib k}'-{\mib q}}a_{{\mib k}}a_{{\mib k}'}, 
\label{eq:Hamiltonian}
\end{eqnarray}
where $\varepsilon_{{\mib k}}$ is the kinetic energy determined by the curvature of the dispersion around the lowest excitation, $\mu$ the chemical potential given by $\mu=g{\mu}_{\rm B}(H-H_{\rm c})$, $U$ the interaction constant and $N$ the number of dimers. If a parabolic isotropic dispersion relation $\varepsilon_{{\mib k}}=({\hbar}{\mib{k}})^2/{2m}$ is used, then the critical chemical potential is given by 
\begin{eqnarray}
{\mu}_{\rm c}=5.22\,U\left(\frac{mk_{\rm B}T}{2{\pi}{\hbar}^2}\right)^{3/2}.
\label{eq:mu}
\end{eqnarray}
This relation leads to the phase boundary described by the power law 
\begin{eqnarray}
\left( g/2 \right) \left[H_{\rm N}(T)-H_{\rm c}\right] = A\,T^{\phi}, 
\label{eq:powerlaw}
\end{eqnarray}
with critical exponent $\phi_{\mathrm{BEC}}=3/2$\,\cite{Nikuni}, where $H_{\rm N}(T)$ is the transition field at temperature $T$. A point given by $T=0$ and $H=H_c$ on the temperature vs field diagram denotes the quantum critical point (QCP). Equation (2) or (3) gives the critical behavior near the QCP characteristic of the magnon BEC. The BEC of magnons has been studied extensively in many gapped spin systems\,\cite{Rueegg,O_KCuCl3,Waki,Tsujii,Vyaselev,Sebastian,Zapf,Stone} and the power law behavior of the phase boundary was confirmed. 
 
TlCuCl$_3$ is an $S = 1/2$ interacting spin dimer system in which a chemical dimer Cu$_2$Cl$_6$ forms a antiferromagnetic spin dimer. The interactions between neighboring dimers are three-dimensional. The lowest excitation occurs at $Q=(0, 0, 1)$ and its equivalent reciprocal points\,\cite{Cavadini2,Oosawa2,Matsumoto}. The magnitude of the excitation gap is ${\Delta}/k_{\rm B} = 7.5$ K\,\cite{Shiramura,O_mag}. 
In the previous magnetization and specific heat measurements in magnetic fields on TlCuCl$_3$\,\cite{O_mag,O_heat}, it was shown that the phase boundary for the FIMO is expressed by the power law with critical exponent $\phi=2.0 \sim 2.2$. This critical exponent is somewhat larger than $\phi_{\mathrm{BEC}} = 3/2$. However, in a lower temperature region $0.5 \rm{K} \leq T \leq 3 \rm{K}$ Shindo and Tanaka obtained rather smaller value $\phi = 1.67$ by the specific heat measurement in magnetic fields\,\cite{Shindo}.

Recently, the deviation of $\phi$ toward larger value from $\phi_{\mathrm{BEC}} = 3/2$ has been discussed theoretically\,\cite{Nohadani,Sherman,Kawashima,MO}. 
Using stochastic series expansion quantum Monte Carlo simulations, Nohadani {\it et al.}\cite{Nohadani} studied the FIMO in cubic lattices of dimers with antiferromagnetic Heisenberg interactions between dimers. They showed that $\phi$ decreases with decreasing fitting temperature range and converges at $\phi_\mathrm{BEC} = 3/2$, and that when fitting range is small enough, $\phi$ is independent of details of the system. They ascribed this behavior to temperature-driven renormalization of the quasiparticle effective mass and of the effective chemical potential. 
On the other hand, Sherman {\it et al.} \cite{Sherman} argued the critical exponent $\phi$, assuming the relativistic dispersion of the form $\varepsilon_{{\mib k}}= \sqrt{{\Delta}^2+C{\mib k}^2}$, because the real dispersion curve around the lowest excitation is better described by the relativistic form than the parabolic one. They showed that the deviation of the dispersion curve from the parabolic form gives rise to the experimental larger exponent, and that the exponent converges at $\phi_\mathrm {BEC} = 3/2$ with decreasing fitting range. Kawashima\,\cite{Kawashima} demonstrated analytically and numerically that the critical exponent $\phi_{\rm BEC} = 3/2$ derived by the HF approximation is exact, although the HF result is usually incorrect for the critical behavior.
Misguich and Oshikawa\,\cite{MO} extended the HF calculation by Nikuni {\it et al.}\,\cite{Nikuni}, using a realistic dispersion relation\,\cite{Cavadini2,Matsumoto} and achieved remarkable quantitative agreement with the experimental phase diagram. They also calculated the critical density of magnons $n_\mathrm{cr}$, which corresponds to the absolute value of the magnetization at $T_\mathrm{N}$.

To the best of the authors' knowledge, TlCuCl$_3$ is the best system to study the magnon BEC and the critical behavior, because details of the magnetic excitations are known and the BEC state can be reached using conventional superconducting magnet. To obtain precise critical exponent $\phi$ and critical density $n_\mathrm{cr}$, we carried out magnetization measurements on TlCuCl$_3$ down to 77 mK.

%\section{Experimental}
Single crystals of TlCuCl$_3$ were grown by the vertical Bridgman method. The  details of preparation were reported in Ref. \citen{O_mag}.
The magnetization measurements were performed using SQUID magnetometer (Quantum Design MPMS XL) in the temperature region $1.8 \mathrm{K} \leq T \leq 100 \mathrm{K}$ in magnetic fields of up to 7 T. The magnetic fields were applied parallel to the $b$-axis and [2, 0, 1] direction and perpendicular to the (1, 0, $\overline{2}$) plane.
The magnetization measurements were also performed using Faraday Force Magnetometer\,\cite{Faraday} at Institute of Solid State Physics in the temperature region $77\  \mathrm{mK} \leq T \leq 4$ K in magnetic fields up to 7 T using a dilution refrigerator. The magnetic field was applied perpendicular to the (1, 0, $\bar{2}$) plane. The gradient field of 5 and 8 T/m was applied to produce the Faraday force.

%\section{Results and Discussion}

\begin{figure}[t]
\includegraphics[scale =0.63]{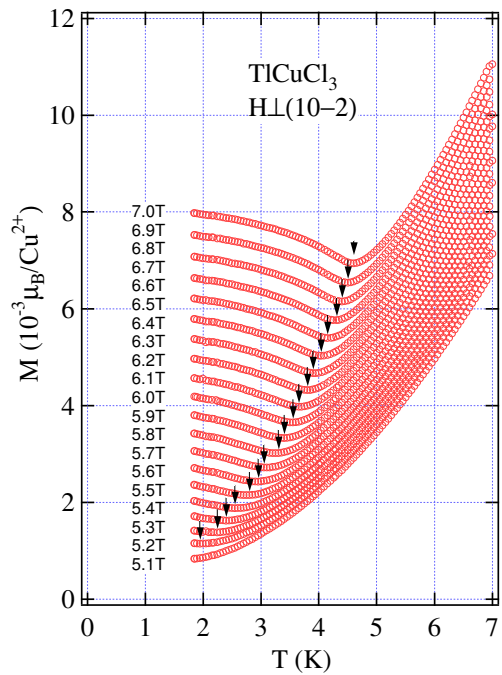}
\caption{(Color online) The temperature dependence of the magnetizations of TlCuCl$_3$ measured at various magnetic fields for $H \bot  (1, 0,  \overline{2})$.}
 \label{fig:mt}
 \end{figure}
Figure \ref{fig:mt} shows low-temperature magnetization $M$ measured at various external fields for $H \perp (1, 0, {\bar 2})$. As temperature is decreased, the magnetization decreases with convex function of temperature and then increases exhibiting the cusplike minimum indicative of the three-dimensional magnetic ordering. This magnetization behavior was observed, irrespective of field direction. We assign the temperature giving magnetization minimum to the ordering temperature $T_{\rm N}$. For $T > T_{\rm N}$, the magnetization corresponds to the number of thermally excited magnons, which decreases with decreasing temperature. The BEC of magnons occurs at $T=T_{\rm N}$, and below $T_{\rm N}$, the number of condensed magnons increases with lowering temperature. The increase of the condensed magnons surpasses the decrease of thermally excited magnons. For this reason, the magnetization has the minimum at $T_{\rm N}$. With decreasing magnetic field, $T_{\rm N}$ decreases, and the cusplike anomaly due to the phase transition becomes smaller. This is because the number of magnons relevant to the BEC decreases when the magnetic field approaches the critical field $H_{\rm c}$.
 
\begin{figure}[t]
\includegraphics[scale =0.63]{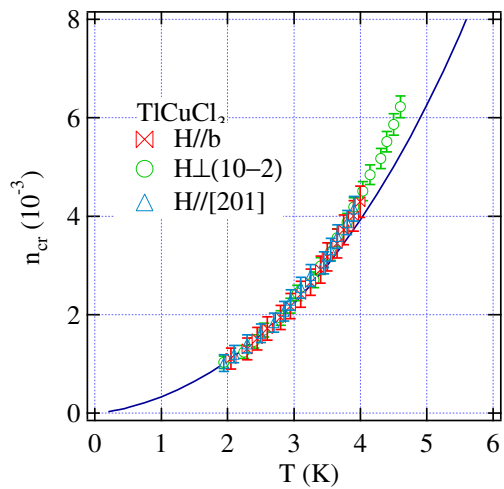}
\caption{(Color online) Temperature dependence of critical density of magnons in TlCuCl$_3$ for $H \| b$, $H \| [2, 0, 1]$ and $H \bot  (1, 0,  \overline{2})$. The solid line is the theoretical calculation by Misguich and Oshikawa\,\cite{MO}.}
 \label{fig:nc}
 \end{figure}

 \begin{figure}[t]
\includegraphics[scale =0.63]{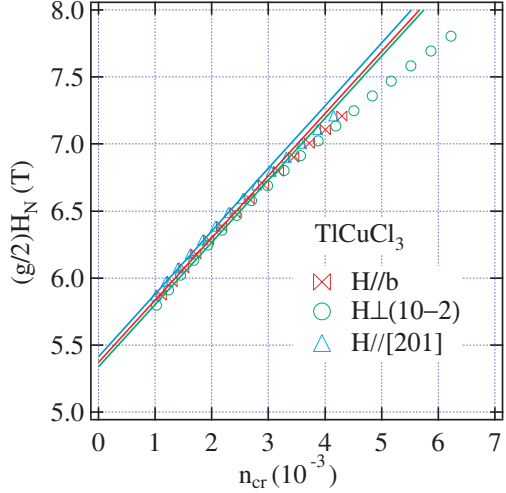}
\caption{(Color online) Transition field $H_{\rm N}$ vs critical density $n_{\rm cr}$ obtained for $H \parallel b$, $H \parallel [2, 0, 1]$ and $H \perp  (1, 0, \overline{2})$. Solid lines are fits by eq. (\ref{eq:U}).}
\label{fig:U}
\end{figure}
The critical density $n_{\rm cr}$ corresponds to the absolute values of the magnetization at $T_\mathrm{N}$ and is obtained from the minimum of the magnetization $M_{\rm cr}$ as $n_{\rm cr}=(2/g{\mu}_{\rm B})M_{\rm cr}$. We corrected the magnetization for the paramagnetism due to impurities which produce the magnetization described by the Brillouin function of $H/T$, and also for the Van Vleck paramagnetism and the diamagnetism due to core electrons. The $g$-factors used are $g = 2.06$ for $H \parallel b$ and $H \parallel [2, 0, 1]$ and $g = 2.23$ for $H \perp (1, 0, {\bar 2})$, which were determined by ESR measurements.
Figure \ref{fig:nc} shows the critical density $n_{\rm cr}$ as a function of temperature. The solid line in Fig. \ref{fig:nc} is the critical density calculated by Misguich and Oshikawa\,\cite{MO} using the realistic dispersion\,\cite{Cavadini2,Matsumoto}. The experimental and theoretical critical densities agree well in the low temperature region of $T \leq 3$ K. However, in the high temperature region, the experimental values are larger than the theoretical ones. The disagreement increases with increasing temperature.

In the HF approximation, the transition field $H_{\rm N} (T)$ is expressed as\,\cite{MO}  
\begin{eqnarray}
(g/2) [H_{\rm N}(T) - H_{\rm c}] = 2Un_\mathrm{cr}(T). \label{eq:U}
\end{eqnarray}
Using eq. (\ref{eq:U}) and the experimental critical density shown in Fig. \ref{fig:nc}, we evaluate the magnon-magnon interaction constant $U$.
Figure \ref{fig:U} shows the plots of transition field $H_{\rm N}(T)$ vs critical density $n_{\rm cr}(T)$, where $H_{\rm N}(T)$ is normalized by the $g$-factor. The linear relation between $H_{\rm N}(T)$ and $n_{\rm cr}(T)$ holds for $n_{\rm cr}(T) \leq 3\times 10^{-3}$. Fitting eq. (\ref{eq:U}) in this small $n_{\rm cr}(T)$ region, we obtain $U/k_\mathrm{B} = 312$ K, 311 K and 315 K for $H \parallel b$, $H \perp  (1, 0,  \overline{2})$ and $H \parallel [2, 0, 1]$, respectively. Their average is $U/k_\mathrm{B} = 313$ K. This value of $U$ is somewhat smaller than $U/k_\mathrm{B} = 340$ K obtained by Misguich and Oshikawa\,\cite{MO} using the magnetization data reported in Ref. \citen{O_mag}. 

\begin{figure}[t]
\includegraphics[scale =0.4]{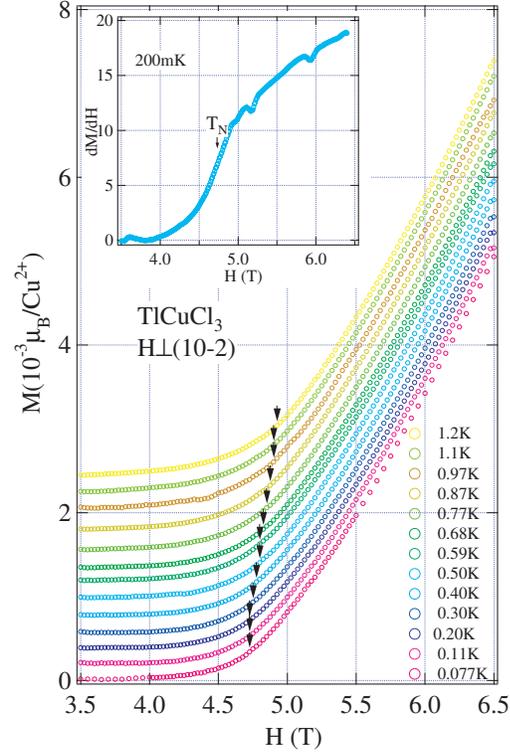}
\caption{(Color online) Magnetization curves in TlCuCl$_3$ measured by Faraday Force Magnetometer for $H \perp  (1, 0,  \overline{2})$ at various temperatures. Backgrounds due to impurities were subtracted. Each plot is shifted vertically by $2\times10^{-4}\mu_{\rm B}$ for clarity. Arrows denote transition fields $H_\mathrm{N}(T)$.}
 \label{fig:faraday}
 \end{figure}
To determine the transition fields for $T < 1.8$ K, we performed magnetization measurements using the Faraday Force Magnetometer.  
Fig. \ref{fig:faraday} shows the magnetization curve measured between 77 mK and 1.2 K for $H \perp (1, 0, \overline{2})$. The raw magnetization has the small background in the low field region due to impurity phase which should be hydrate on the sample surface. In Fig. \ref{fig:faraday}, the background magnetization was corrected. The magnetization is almost zero up to the transition field $H_\mathrm{N}\simeq 4.7$ T and increases rapidly. The gapless ordered state appears for $H > H_\mathrm{N}$. The magnetization does not show sharp bend at $H_\mathrm{N}$, but is rather rounded even at 77 mK. This is ascribed not to the gradient field to produce the Faraday force, because the distribution of magnetic field in the sample is less than 0.04 T. We infer that the antisymmetric interaction of the Dzyaloshinsky-Moriya type which mixes the singlet and triplet states gives rise to the smearing of the magnetization anomaly at $H_\mathrm{N}$. We assign the transition field $H_{\rm N}(T)$ to the field of inflection in the derivative of magnetization $dM/dH$, as shown in the inset of Fig. \ref{fig:faraday}. 

\begin{figure}[t]
\includegraphics[scale =0.55]{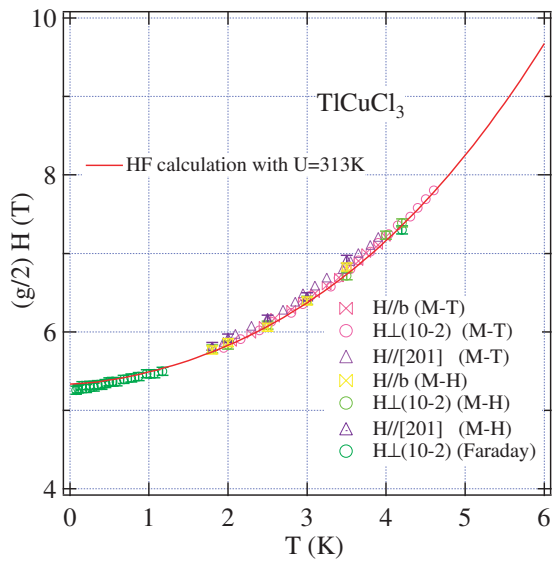}
\caption{(Color online) Phase diagram normalized by the $g$-factors in TlCuCl$_3$. The solid line is the result of calculation using the realistic dispersion and the interaction constant of $U/k_{\rm B} = 313$ K. \cite{MO} }
 \label{fig:phase}
 \end{figure}

The phase transition points obtained by temperature and field scans of magnetization are summarized in Fig. \ref{fig:phase}. Since the phase boundaries for $H \parallel b$, $H \parallel [2, 0, 1]$ and $H \perp  (1, 0,  \overline{2})$ coincide when normalized by the $g$-factor, we can deduce that the phase boundary is independent of the external field direction, which implies that the magnetic anisotropy is negligible in TlCuCl$_3$. Solid line in Fig. \ref{fig:phase} is the HF calculation using the realistic dispersion\,\cite{MO,Cavadini2,Matsumoto} and the interaction constant of $U/k_{\rm B} = 313$ K obtained by our magnetization measurement. The experimental phase boundary for $T < 5$ K agrees almost perfectly with the calculation.

\begin{figure}[t]
\includegraphics[scale =0.55]{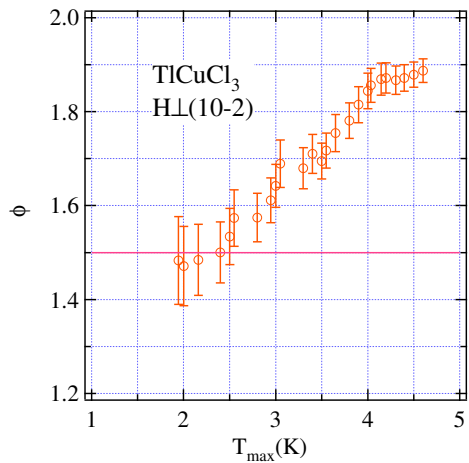}
\caption{(Color online) Critical exponent $\phi$ as a function of $T_\mathrm{max}$ obtained by fitting eq. (\ref{eq:powerlaw}) to transition points between 77 mK and $T_\mathrm{max}$ (K). The transition points were collected for $H \perp (1,0,\bar 2)$}
\label{fig:phi}
\end{figure}

We analyze the phase boundary for the magnetic field perpendicular to the $(1,0,\bar 2)$ plane using the power law given by eq. (\ref{eq:powerlaw}). We fit eq. (\ref{eq:powerlaw}) in the temperature range of $T_{\rm min} \leq T \leq T_{\rm max}$, setting the lowest temperature at $T_{\rm min}=77$ mK and varying the highest temperature $T_{\rm max}$ from 4.6 K to 1.94 K. The critical exponent $\phi$ obtained by the fit is graphed in Fig. \ref{fig:phi}. It is observed that the critical exponent $\phi$ decreases with decreasing $T_\mathrm{max}$, and converges at $\phi_\mathrm{BEC} = 3/2$, as recent theory predicts \cite{Nohadani,Sherman,MO}. For $T_\mathrm{max} \leq 2.4$ K, we obtain $(g/2)H_{\rm c} = 5.27 \pm 0.01$ T, $A = 0.225 \pm 0.01$ T/K$^{\phi}$ and $\phi = 1.50 \pm 0.06$. When ${\phi}=3/2$, the parameter $A$ corresponds to the coefficient of $T^{3/2}$ in eq. (\ref{eq:mu}), and is given by
\begin{eqnarray}
A = \frac{2.61\,U}{\mu_B}\left(\frac{mk_B}{2\pi \hbar^2}\right)^{\frac{3}{2}}. \label{eq:A}
\end{eqnarray}
Substituting $A = 0.225$ T/K$^{1.5}$ into eq. (\ref{eq:A}), we obtain $mk_{\rm B}/{\hbar}^2=0.0204$ K$^{-1}$. This value is consistent with $mk_{\rm B}/{\hbar}^2=0.0229$ K$^{-1}$ evaluated from the curvature of the dispersion relation around the lowest excitation \cite{MO}. In eq. (\ref{eq:mu}), the lattice constant is chosen to be unity. If we use an average lattice constant ${\bar a}$ given by ${\bar a}=(abc\sin{\beta})^{1/3}$, where $a=3.98$ \AA, $b=14.14$ \AA, $c=8.89$ \AA and ${\beta}=96.32^{\circ}$ are lattice parameters in TlCuCl$_3$ \cite{Tanaka}, we obtain $m=2.61\times 10^{-26}$ g. This effective magnon mass is approximately $1/60$ of the proton mass. This smaller magnon mass enables the BEC at helium temperatures in spite of small density of order of $10^{-3}$.

%\section{Conclusion}
In conclusion, we have presented the results of magnetization measurements performed on TlCuCl$_3$ at temperatures down to 77 mK and in magnetic fields up to 7 T for three different field directions. The critical density of magnons as a function of temperature and the magnetic phase diagram for external field vs temperature were obtained. The magnon-magnon interaction constant was estimated as $U/k_B = 313$ K. The phase boundary is expressed by the power law of eq. (\ref{eq:powerlaw}) and agrees almost perfectly with the magnon BEC theory based on the Hartree-Fock approximation with realistic dispersion relations \cite{MO} and $U/k_B = 313$ K obtained by our magnetization measurement. The critical exponent ${\phi}$ decreases with decreasing fitting temperature range. For $77\,\mathrm{mK} \leq T \leq 2.4$ K, we obtained $\phi = 1.50 \pm 0.06$, which coincides with $\phi_\mathrm{BEC} = 3/2$ derived from the magnon BEC theory \cite{Nikuni,Kawashima}. The effective magnon mass obtained by the present analysis is consistent with that evaluated from the curvature of the dispersion relation around the lowest excitation. The present results strongly support the BEC description of the field-induced magnetic ordering in TlCuCl$_3$. 

%\section*{Acknowledgment}
The authors would like to thank  M. Matsumoto for showing us their theoretical calculations and stimulating discussions. This work was supported by a Grant-in-Aid for Scientific Research from the Japan Society for the Promotion of Science and by a 21st Century COE Program at Tokyo Tech ``Nanometer-Scale Quantum Physics'' from the Ministry of Education, Culture, Sports, Science and Technology.

\end{document}